\documentclass[aps,pre,reprint,superscriptaddress,floatfix]{revtex4-2}

\usepackage{amsmath,amssymb,amsfonts,mathtools}
\usepackage{graphicx}
\usepackage{xcolor}
\usepackage{makecell}
\usepackage{tikz}
\usepackage{pgfplots}
\pgfplotsset{compat=newest}
\usepackage{hyperref}

\newcommand{\e}{\mathrm{e}}
\newcommand{\sgn}{\,\text{sgn}}
\newcommand*\diff{\mathop{}\!\mathrm{d}}

\newcommand{\eqn}[1]{Eq.~(\ref{#1})}
\newcommand{\Eqn}[1]{Equation~(\ref{#1})}

\begin{document}

\title{Equilibrating continuous-variable open quantum systems using stochastic classical trajectories in path-integral space}

\author{William H. D. Moore}
\author{Stuart C. Althorpe}
\affiliation{Yusuf Hamied Department of Chemistry, University of Cambridge, Lensfield Road, Cambridge, CB2 1EW, United Kingdom}

\date{\today}

\begin{abstract}
    Beyond the weak-coupling limit, open quantum systems equilibrate to a highly entangled thermal state. For continuous-variable systems, this state can be written explicitly as an imaginary-time phase-space path integral, in which the positions are directly entangled with the bath, and the momenta are correlated with the positions through a phase term. Here, we ask to what extent this state can be reached by propagating stochastic classical trajectories in path-integral phase space. Surprisingly, we find that the trajectories equilibrate to the exact quantum equilibrium state, recovering the purely imaginary momentum--position correlation in the phase term. The trajectories are generated using a recently derived Matsubara generalized Langevin equation, which produces the imaginary correlations by evolving the stochastic variables into the complex plane. This makes the dynamics numerically unstable, but we are nonetheless able to demonstrate the equilibration of a quartic oscillator coupled to a white-noise bath. These unexpected findings could lead to new approximate methodologies for simulating continuous-variable open quantum systems.
    \end{abstract}

\maketitle

\section{Introduction}\label{sec:intro}

Continuous-variable open quantum systems are used to model the dynamics of a wide variety of processes, including quantum Brownian motion \cite{Caldeira1983,Grabert1988}, surface diffusion \cite{miret2005dynamics, TorresMiyares2022,  Trenins2025} and vibrational spectroscopy \cite{Tanimura1993, Kato2004}. The total Hamiltonian is usually written in the form \cite{Caldeira1983,weiss2012quantum}
\begin{align}\label{1}
\hat H = \hat H_\text{s} +  \hat H_\text{b} +  \hat H_\text{sb}
\end{align}
where the system Hamiltonian
\begin{align}\label{2}
\hat H_\text{s}={\hat p^2\over2m} + V(\hat q)
\end{align}
is a continuous function of $\hat p$ and $\hat q$, the bath Hamiltonian $H_\text{b}$ is a collection of harmonic oscillators, and the coupling term $\hat H_\text{sb}=\hat q\hat F$, where $\hat F$ is the collective bath/solvent coordinate.

For an open quantum system, a non-equilibrium reduced density matrix (obtained by tracing the full density matrix over the locally equilibrated bath states) will equilibrate to the fully entangled equilibrium state after propagating for long enough with the exact quantum time-evolution operator. For a continuous-variable system, this equilibrium state can be written out explicitly in terms of imaginary-time path-integral phase-space variables \cite{weiss2012quantum,Althorpe2021,prada2023comparison}. The entanglement appears as a Gaussian term  reducing the amplitude of  thermal fluctuations around the path centroids in response to the bath coupling. Crucially, the system momenta and positions are correlated through a phase term.

In this article we investigate whether it is possible to equilibrate to this state by propagating stochastic classical trajectories in path-integral phase space. It is well known that this can be done for the position marginal, using the well-established technique of path-integral molecular dynamics, whereby  a thermostat is attached to a fictitious `ring-polymer' Hamiltonian \cite{Parrinello1984,Lawrence2019}. However, generating the full momentum--position distribution from classical trajectories would appear to be impossible on account of the phase, which stochastic classical dynamics cannot be expected to generate. Surprisingly, we show in this article that this is not the case: stochastic classical trajectories can, at least in principle, and sometimes in practice (depending on numerical stability), equilibrate to the exact quantum equilibrium state.

The key is to generate the trajectories using a generalised Langevin equation (GLE) in path-integral phase-space, which was derived recently in Ref.~\onlinecite{prada2023comparison} using Matsubara dynamics. `Matsubara dynamics' in this context refers to a semiclassical approximation to the exact quantum dynamics for a continuous system, which arises when initially smooth imaginary-time Feynman paths are assumed to remain smooth for all time \cite{Althorpe2021,hele2015boltzmann}. This sole approximation is sufficient to collapse the exact quantum dynamics to an ensemble of classical trajectories in the extended path-integral phase space. Ref.~\onlinecite{prada2023comparison} showed that, when applied to a system--bath Hamiltonian of the form of \eqn{1}, Matsubara dynamics can be used to derive a GLE in the path-integral phase space of the system. 

In Ref.~\onlinecite{prada2023comparison} it was assumed that, although this GLE conserves the quantum Boltzmann distribution, it would not be capable of taking an arbitrary initial distribution and equilibrating it, on account of the momentum--position phase \footnote{To compensate for this, Ref.~\onlinecite{prada2023comparison} included the momentum--position correlation explicitly in the reduced density matrix at time $t=0$ using analytic continuation.}. Here, we show that it does precisely this, by evolving the system into the complex plane such that the resulting distribution naturally includes the phase and correctly describes the imaginary correlations between the momenta and positions. To prove that the Matsubara GLE has this property, we map it onto an auxiliary space in which the stochastic dynamics becomes Markovian, from which we construct a Fokker--Planck equation, the stationary solution of which is then shown to marginalise to the exact quantum equilibrium state. 
 
The article is structured as follows. Sections~\ref{sec:open_classical_systems}--\ref{sec:oqs_matsubara} summarise key results from prior work on classical open systems, and path-integral descriptions of closed and open quantum systems respectively, with  Sec.~\ref{sec:oqs_matsubara_modified_GLE} reporting a slightly modified version of the Matsubara GLE of Ref.~\onlinecite{prada2023comparison}. Most of the new work, including mapping the Matsubara GLE onto an auxiliary space, deriving the Matsubara Fokker--Planck equations, demonstrating that the stationary solutions marginalise to the thermal equilibrium state, and reporting a simple numerical confirmation of these theoretical predictions, is reported in Sec.~\ref{sec:exact_quatnum_thermostat}. Sec.~\ref{sec:conc} concludes the article.

\section{Classical background}\label{sec:open_classical_systems}

\subsection{Generalised Langevin equations}
Consider an open classical system with the usual system--bath Hamiltonian \cite{weiss2012quantum,tuckerman2010statistical}
\begin{multline}\label{eq:caldeira_leggett_classical}
    H(p,q, \boldsymbol{p},\boldsymbol{x}) =  H_\text{s}(p,q)+ \sum_{\alpha=1}^{n_{b}}\Bigg[ \frac{p_{\alpha}^{2}}{2m_{\alpha}} \\
    + \frac{1}{2} m_{\alpha} \omega_{\alpha}^{2} \left( x_{\alpha}  - \frac{c_{\alpha}}{m_{\alpha}\omega_{\alpha}^{2}} q \right)^{2}\Bigg]
\end{multline}
where
\begin{equation}
H_\text{s}(p,q) = \frac{p^{2}}{2m} + V(q)
\end{equation}
and the coefficients $c_{\alpha}$ are generated from a spectral density
\begin{equation}\label{eq:spectral_density_definition}
    J(\omega) = \frac{\pi}{2}\sum_{\alpha=1}^{n_{b}}\frac{c_{\alpha}^{2}}{m_{\alpha}\omega_{\alpha}}\delta(\omega-\omega_{\alpha}).
\end{equation}
The dynamics of the system coordinates $(p,q)$ can be propagated using a generalised Langevin equation (GLE) 
\begin{equation}\label{eq:GLE_classical}
    m\ddot{q}(t) = -V'[q(t)] - \int_{0}^{t}\!\diff s\, \zeta(t-s)\dot{q}(s) + R(t)
\end{equation}
in which 
 \begin{equation}\label{eq:random_force_classical_randomsampling}
    R(t) = \sum_{\alpha=1}^{n_{b}}\sqrt{\frac{c_{\alpha}^{2}}{\beta m_{\alpha}\omega_{\alpha}^{2}}}\left[ \lambda_{\alpha}\cos{\omega_{\alpha}t} + \xi_{\alpha}\sin{\omega_{\alpha}t} \right]
\end{equation}
where $\beta=1/k_\text{B}T$, $\lambda_{\alpha}$ and $\xi_{\alpha}$ are Gaussian random variates with zero mean and unit variance, and \begin{align}\label{eq:kernelfromJintegral}
    \zeta(t-s) &= \beta\langle R(s)R(t)\rangle \nonumber\\
    &= \sum_{\alpha=1}^{n_{b}}\frac{c_{\alpha}^{2}}{m_{\alpha}\omega_{\alpha}^{2}}\cos{\left(\omega_{\alpha}(t-s)\right)}\nonumber\\
    &= \frac{2}{\pi}\int_{0}^{\infty}\!\diff\omega\,\frac{J(\omega)}{\omega}\cos\left(\omega (t-s)\right).
\end{align}
This equation is the fluctuation--dissipation relation \cite{kubo1966fluctuation} that ensures that an arbitrary, normalised, system probability density evolves to the equilibrium distribution $Z^{-1}\exp[-\beta H_\text{s}(p,q)]$ as $t\to\infty$.

In the special case of an ohmic spectral density
\begin{equation}\label{eq:white_spectral_density}
    J(\omega)=m\gamma\omega
\end{equation}
the noise $R(t)$ becomes white (and will then be written $W(t)$), such that
 \begin{equation}\label{eq:kernel_white}
    \zeta(t-s) = \beta\langle W(s)W(t)\rangle = 2m\gamma \, \delta(t-s)
\end{equation}
and the dynamics becomes Markovian, with  \eqn{eq:GLE_classical} reducing to the Langevin equation
\begin{equation}\label{eq:LE_classical}
    m\ddot q(t) = -V'[q(t)] - m\gamma \dot q(t) + W(t).
\end{equation}

\subsection{Fokker--Planck equations for Markovian dynamics}

A stochastic, Markovian, equation of motion such as \eqn{eq:LE_classical} is equivalent to a deterministic  Fokker--Planck equation (FPE) describing the time-evolution of the probability density $\rho(p,q, t)$. There is a standard procedure for generating the FPE \cite{zwanzig2001nonequilibrium,risken1989fokker}. The stochastic equations of motion are put into the form
\begin{equation}\label{eq:general_eom_continuous}
    \dot{\mathbf{x}}(t)=\mathbf{G}(\mathbf{x}(t)) + \mathbf{F}(t)
\end{equation}
where  $\mathbf{x}(t)$ is a vector composed of the $n$ dynamical variables,  
and
\begin{equation}
    F_{i}(t)=\sum_{j=1}^{n}\alpha_{ij}\eta_{j}(t)
\end{equation}
in which $\eta_j(t)$ are white noise-variables with zero mean and unit variance.
The FPE corresponding to \eqn{eq:general_eom_continuous} can then be shown to be \cite{zwanzig2001nonequilibrium}
\begin{equation}\label{eq:FP_general}
    \frac{\partial \rho(\mathbf{x}, t)}{\partial t} = \left[-\frac{\partial}{\partial \mathbf{x}}\cdot\mathbf{G}(\mathbf{x}) + \frac{\partial}{\partial \mathbf{x}}\cdot\left(\mathbf{\Theta}\frac{\partial}{\partial \mathbf{x}}\right)\right]\rho(\mathbf{x}, t)
\end{equation}
where
\begin{equation}\label{eq:theta_matrix_elements}
    \Theta_{ij} = \frac{1}{2}\sum_{k=1}^{n}\alpha_{ik}\alpha_{jk}.
\end{equation}

Applying this procedure to \eqn{eq:LE_classical} yields the well-known Klein--Kramers equation
\begin{multline}\label{eq:FP_white_classical}
    \frac{\partial \rho(p,q,t)}{\partial t} = \bigg[-\frac{\partial}{\partial q}\frac{p}{m} + \frac{\partial}{\partial{p}}\frac{\diff V}{\diff{q}} \\
    + \gamma\frac{\partial}{\partial p}\left(p+\frac{m}{\beta}\frac{\partial}{\partial p}\right)\bigg]\rho(p,q,t).
\end{multline}

\subsection{Fokker--Planck equations for non-Markovian dynamics}\label{sec:classical_auxiliary_variables}

To obtain the FPE corresponding to a non-Markovian stochastic dynamics, we need to map onto an auxiliary space in which the dynamics is Markovian \cite{baczewski2013numerical}. To illustrate this procedure, consider the GLE of \eqn{eq:GLE_classical} with a Debye--Drude spectral density
\begin{equation}\label{eq:general_spectral_density}
    J(\omega) = m\omega\frac{\gamma\omega_{c}^{2}}{\omega^{2}+\omega_{c}^{2}}
\end{equation}
for which the kernel is
\begin{equation}\label{eq:ddnoise}
    \zeta(t-s) = \beta \langle R(s)R(t)\rangle = m\gamma\omega_{c} \e^{-\omega_{c}\left|t-s\right|}.
\end{equation}
The first step is to introduce a stochastic variable $y(t)$, defined to be the 
solution to the Ornstein--Uhlenbeck equation
\begin{equation}\label{eq:OU_process_continuous_time}
    \dot y(t) = -\omega_{c} y(t) + \omega_{c} W(t),
\end{equation}
in which $W(t)$ is the white-noise random variate defined in \eqn{eq:kernel_white}. The solution to this equation for $t>0$ is
 \begin{equation}
    y(t) = y(0) e^{-\omega_{c} t} + \omega_{c}\int_0^t\!\diff s\,e^{-\omega_{c}(t-s)} W(s)
\end{equation}
which gives
\begin{equation}\label{eq:OU_process_correlations}
    \langle y(s)y(t)\rangle = {m\gamma\omega_{c}\over\beta}\e^{-\omega_{c}|t-s|} + \left[ \langle y^{2}(0) \rangle -  {m\gamma\omega_{c}\over\beta}\right]\e^{-\omega_{c}(t+s)}.
\end{equation}
When $y(0)$ is sampled from a Gaussian distribution of variance $m\gamma\omega_{c}/\beta$, the resulting $y(t)$ thus gives $R(t)$ of \eqn{eq:ddnoise}. The second step is to introduce a variable  $v(t)$, defined to be the solution to
\begin{equation}\label{eq:v_eom}
    \dot{v}(t) = -\omega_{c}v(t) + m\gamma \omega_{c}\dot{q}(t)
\end{equation}
which is
\begin{equation}
    v(t) = v(0)\e^{-\omega_{c}t} + m\gamma \omega_{c}\int_{0}^{t}\!\diff s\,\e^{-\omega_{c}(t-s)}\dot{q}(s).
\end{equation}
When $v(0)$ is set to zero, $v(t)$ gives the negative of the friction term in \eqn{eq:GLE_classical}. Combining these two variables into
\begin{equation}
z(t)=y(t)-v(t)
\end{equation}
allows us to write
\begin{align}\label{eq:marko}
    \dot{q}(t)&=\frac{p(t)}{m} \nonumber\\
    \dot{p}(t) &= -\frac{\diff V(q(t))}{\diff q} + z(t) \nonumber\\
    \dot{z}(t)&=-\omega_{c}z(t) - \gamma\omega_{c} p(t) + \omega_{c}W(t)
\end{align}
where the dynamics of $(p,q)$ is identical to that generated by the GLE of \eqn{eq:GLE_classical}.

Since the dynamics of \eqn{eq:marko} is Markovian, it is straightforward to obtain ${\bf G}$ and ${\boldsymbol{\Theta}}$ of \eqn{eq:FP_general}, and thus to construct the FPE in the auxiliary space, which is
\begin{multline}\label{eq:FP_debye_classical}
    \frac{\partial \rho(p,q,z,t)}{\partial t} = \bigg[-\frac{\partial}{\partial q} \frac{p}{m}+ \frac{\partial}{\partial{p}}\left(\frac{\diff V}{\diff{q}}-z\right) \\
    + \gamma\omega_{c}\frac{\partial}{\partial z}p + \omega_{c}\frac{\partial}{\partial z}\left(z+\frac{m\gamma\omega_{c}}{\beta}\frac{\partial}{\partial z}\right)\bigg]\rho(p,q,z,t).
\end{multline}
The stationary solution 
\begin{equation}
\rho_\text{stat}(p,q,z)={1\over \cal{N}} \exp\left[-\beta\left({p^2\over 2m} +V(q) + {z^2\over 2m\gamma\omega_{c}}\right) \right]
\end{equation}
(where $\mathcal{N}$ normalises the distribution here and throughout) marginalises to give $Z^{-1}\exp\left[-\beta\left({p^2/ 2m} +V(q)\right)\right]$ as required. 

With the above procedure in hand, it is straightforward to construct the FPE for any spectral density $J(\omega)$, provided it can be written as a Meier--Tannor expansion \cite{Meier1999} over $N_\text{L}$ Lorentzians. Since each Lorentzian term is equivalent to a Debye--Drude spectral density with complex poles, the auxiliary variables resemble a straightforward $N_\text{L}$-dimensional generalisation of the variable $z(t)$ in \eqn{eq:marko}. For this reason, we will not need to generalise the derivation of Sec.~\ref{sec:exact_quatnum_thermostat} beyond a Debye--Drude spectral density. 

\section{Imaginary-time Path integrals}\label{sec:matsubara_dynamics}

\subsection{Equilibrium distributions}

For an isolated system, with Hamiltonian $\hat H_\text{s}$, the position marginal of the path-integral  equilibrium distribution is \cite{ Parrinello1984,Chandler1981}
\begin{align}\label{eq:pimd}
   \rho_\text{eq}({\bf q})&={1\over Z}\prod_{i=l}^N \langle q_l|e^{-\beta_N\hat H_\text{s}}| q_{l+1} \rangle \nonumber\\
   &={1\over \cal{N}}\e^{-\beta_N\left[U_N({\bf q}) + S_N({\bf q})\right]}
\end{align}
where $\beta_N=\beta/N$, $q_{N+1}\equiv q_1$, and
\begin{subequations}
\begin{align}
U_N({\bf q}) &=\sum_{l=1}^NV(q_l)\\
S_N({\bf q}) &=\sum_{l=1}^N  {(q_{l+1}-q_l)^2m\over 2 \beta_N\hbar^2}.
\end{align}
\end{subequations}
The imaginary-time Feynman paths ${\bf q}\equiv\left\{q_i\right\}_{i=1}^N$ are Wiener paths in imaginary time $\tau=0\to\beta\hbar$ which resemble jagged loops, often nicknamed `ring polymers'. 

The paths can be Fourier-smoothed into differentiable functions of $\tau$ by approximating them as \cite{ceperley1995path}
\begin{equation}\label{eq:four}
    q\left(\tau\right)=Q_{0}+\sqrt{2}\sum_{n=1}^{\widetilde{M}}\left[Q_{n}\sin\omega_{n}\tau + Q_{\bar{n}}\cos\omega_{n}\tau\right]
\end{equation}
with $\widetilde{M} = (M-1)/2$ \footnote{We assume throughout that $M$ is odd.}, where $M$ is the number of Fourier modes, and
 \begin{align}
 \omega_n = {2 n \pi\over \beta\hbar}
\end{align} 
are the Matsubara frequencies. Formally, this smoothing is produced by making $M$ finite, then taking the limit $N\to\infty$ in the distribution of \eqn{eq:pimd}. The mode $Q_0$ is often referred to as the `centroid' and the modes $Q_n,n\ne0$ as the `Matsubara modes' \cite{Althorpe2021,hele2015boltzmann}. The latter describe the quantum thermal fluctuations around the centroid and shrink to zero  in the limit $\beta\to0$. [Note that in the above and in what follows we use the convention that positive/negative values of $n$ denote sin/cos modes (where $\bar n\equiv -n$) and that these signs are included in the definition of $\omega_n$; e.g.\ $\omega_{\bar{2}}\equiv \omega_{-2}=-4\pi/\beta\hbar$.]

On smoothing the paths, \eqn{eq:pimd} becomes
\begin{equation}\label{eq:rhoQ}
   \rho_\text{eq}({\bf Q})={1\over \cal{N}}\e^{-\beta\left[U_M({\bf Q}) + S_M({\bf Q})\right]}
   \end{equation}
where
\begin{subequations}
\begin{align}\label{eq:mats_potential}
    U_{M}(\mathbf{Q}) &= {1\over\beta\hbar}\int_0^{\beta\hbar}\!\diff\tau\, V[q(\tau)]\\
    S_M({\bf Q}) &={m\over 2\beta\hbar}  \int_0^{\beta\hbar}\!\diff\tau\,  \left({\partial q(\tau)\over \partial \tau}\right)^2 \nonumber\\
    &={m\over 2}\sum_{n=-\widetilde M}^{\widetilde M}\omega_n^2 Q_n^2.
\end{align}
\end{subequations}

One can obtain the fully correlated position--momentum distribution from \eqn{eq:rhoQ} by inverting the Fourier transforms over momentum that produced the spring term $S_M({\bf Q})$, to obtain \cite{Althorpe2021}
 \begin{equation}\label{eq:rhoco}
   \rho_\text{eq}({\bf P},{\bf Q})={1\over \cal{N}}\e^{-\beta\left[H_M({\bf P},{\bf Q}) -i\theta_{M}(\mathbf{P}, \mathbf{Q}) \right]}
   \end{equation}
   where
\begin{align}\label{eq:matham}
   H_M({\bf P},{\bf Q}) &={1\over\beta\hbar}\int_0^{\beta\hbar}\!d\tau\, \left({p^2(\tau)\over 2m} +V[q(\tau)]\right)\nonumber\\
   &={{\bf P}^2\over 2m} + U_M({\bf Q})\\
\label{eq:phase}
    \theta_{M}(\mathbf{P}, \mathbf{Q}) &= -{1\over\beta\hbar}\int_0^{\beta\hbar}\!d\tau\, {\partial q(\tau)\over \partial \tau}p(\tau) \nonumber\\
   &= \sum_{n=-\widetilde{M}}^{\widetilde{M}}\omega_{n}Q_{\bar{n}}P_{n} 
\end{align}
and $p(\tau)$ and ${\bf P}\equiv\{P_n \}$ are defined by analogy with \eqn{eq:four}.
We will refer to $H_M({\bf P},{\bf Q})$ as the `Matsubara Hamiltonian' and $\theta_{M}(\mathbf{P}, \mathbf{Q}) $ as the `Matsubara phase'. It is well known \cite{ceperley1995path, coalson1986connection} that the distribution $\rho_\text{eq}({\bf P},{\bf Q})$ yields the exact quantum static expectation values of any operator function of $(\hat p,\hat q)$.

 \subsection{Matsubara dynamics}

Under real-time quantum evolution, a function which initially depends only on the smooth Matsubara paths $p(\tau)$ and $q(\tau)$, or equivalently, the Matsubara modes $({\bf P},{\bf Q})$, rapidly becomes dependent also on the jagged, discontinuous modes that have been excluded from \eqn{eq:four}---unless $V(q)$ is harmonic. This is because the exact time evolution is generated by applying $N$ replicas of the von-Neumann--Liouville operator, one at each of the discrete imaginary time-slices in \eqn{eq:pimd}. However, it is possible to project the jagged modes out of this operator, to ensure that an initially smooth function remains smooth for all time \cite{Althorpe2021,hele2015boltzmann}. The application of this constraint makes the dynamics classical in the extended phase space $({\bf P},{\bf Q})$, where the trajectories follow the equations of motion
\begin{align}\label{eq:matty}
\dot Q_n & = {\partial H_M({\bf P},{\bf Q})\over \partial P_n}={P_n \over m} \nonumber\\
\dot P_n & =-{\partial H_M({\bf P},{\bf Q})\over \partial Q_n}= -{\partial U_M({\bf Q})\over \partial Q_n}
\end{align}
in which $H_M({\bf P},{\bf Q})$ is the Matsubara Hamiltonian of \eqn{eq:matham}. We will refer to this approximate dynamics as `Matsubara dynamics'.

Despite being classical, Matsubara dynamics conserves the quantum Boltzmann distribution of \eqn{eq:rhoco}. This is because $H_M({\bf P},{\bf Q})$ is conserved from \eqn{eq:matty}, and $\theta_M({\bf P},{\bf Q})$ is conserved because it is the momentum conjugate to a continuous symmetry transformation
\begin{equation}
\tau\to\tau+\tau_0
\end{equation}
under which $H_M({\bf P},{\bf Q})$ is invariant. This last property follows from the periodicity of the imaginary-time paths, which allows one to write
    \begin{equation}
    \frac{\partial H_{M}}{\partial\tau_{0}} = 0.
    \end{equation}
 Additional properties of this imaginary-time-translation symmetry are given in Appendix~\ref{app:immaginary_time_symmetries}. Two key relations that will be used later in Sec.~\ref{sec:exact_quatnum_thermostat} are
\begin{align}\label{vint}
    \frac{\partial U_{M}}{\partial\tau_{0}}    &=- \sum_{n=-\widetilde{M}}^{\widetilde{M}} \omega_{n}Q_{\bar{n}} \frac{\partial U_{M}}{\partial Q_{n}}=0
\end{align}
and
\begin{align}\label{eq:mats_mode_symmetry_1}
    R_{n}\left(\tau_{0} + \frac{\beta\hbar}{4n}\right) &= -R_{\bar{n}}\left(\tau_{0}\right)\nonumber\\
    R_{\bar{n}}\left(\tau_{0} + \frac{\beta\hbar}{4n}\right) &= R_{n}\left(\tau_{0}\right)
\end{align}
where $R$ denotes $P$ or $Q$.

\section{Path-integral description of open quantum systems}\label{sec:oqs_matsubara}

The results given in the previous section generalise straightforwardly to the full Caldeira--Leggett Hamiltonian 
$\hat H$ of \eqn{1}. Here we summarise the reduced picture in the system coordinates, obtained by tracing over the bath modes. 

\subsection{Reduced equilibrium distribution}

The reduced phase-space equilibrium distribution is \cite{prada2023comparison,moore2026matsubara}
\begin{equation}\label{eq:redeq}
   \rho_\text{eq}({\bf P},{\bf Q})={1\over \cal{N}}\e^{-\beta\left[F_M({\bf P},{\bf Q}) -i\theta_{M}(\mathbf{P}, \mathbf{Q}) \right]}
   \end{equation}
where $({\bf P},{\bf Q})$ are the Matsubara modes of the system,
\begin{equation}\label{eq:fm}
F_M({\bf P},{\bf Q}) =H_M({\bf P},{\bf Q}) + \frac{m}{2}\sum_{n=-\widetilde M}^{\widetilde M}\left|\omega_{n}\right| \widetilde{\zeta}(\left|\omega_{n}\right|)Q_{n}^{2}
 \end{equation}
and $H_M({\bf P},{\bf Q})$ is given by \eqn{eq:matham}, with $V(q)$ taken to be the system potential of \eqn{2}. The phase $\theta_{M}(\mathbf{P}, \mathbf{Q})$  is now the system contribution to the total Matsubara phase, the bath contribution having been integrated out to give the second term in \eqn{eq:fm} which describes the system--bath entanglement. The term $\widetilde{\zeta}(s)$ is given by
   \begin{align}\label{eq:zeta_hat}
    \widetilde{\zeta}(s) &= \sum_{\alpha=1}^{n_{b}}\frac{c_{\alpha}^{2}}{ m_{\alpha}\omega_{\alpha}^{2}}\frac{s}{s^{2}+\omega_{\alpha}^{2}}\nonumber\\
    &= \int_0^\infty\!\diff t\,\zeta(t) e^{-st}.
\end{align}
For the white-noise spectral density of \eqn{eq:white_spectral_density}, $\widetilde{\zeta}(|\omega_n|)=\gamma|\omega_n|$, which illustrates clearly that increasing the system--bath coupling strength $\gamma$ decreases the variance of the quantum thermal fluctuations  around the centroid; this trend also holds for other spectral densities.

\subsection{Matsubara generalised Langevin equation}\label{sec:oqs_matsubara_modified_GLE}

Reference \onlinecite{prada2023comparison} reported a GLE which was derived by starting from the Matsubara dynamics of the system plus bath, then eliminating the bath modes. This derivation assumed a direct product initial condition. Appendix~\ref{app:gle_derivation} of this article reports a modified derivation, which assumes an initial distribution in which the bath is locally equilibrated around the system coordinate. The resulting GLE is
\begin{multline}\label{eq:GLE_matsubara}
    m\ddot{Q}_{n}(t) = f^{(n)}_{M}(\mathbf{Q}(t)) - \int_{0}^{t}\!\diff s\, \zeta(t-s)\dot{Q}_{n}(s) + R_{n}^{\mathbb{C}}(t) \\
      - \left[Q_{n}(0)K_{n}(t) - iQ_{\bar{n}}(0)L_{n}(t) \right]
\end{multline}
where
\begin{align}\label{eq:force}
f^{(n)}_{M}(\mathbf{Q}) = -{\partial U_M(\mathbf{Q})\over \partial Q_n}
\end{align}
and the complex-valued noise is
\begin{multline}\label{eq:complex_noise}
    R_{n}^{\mathbb{C}}(t) = \sum_{\alpha=1}^{n_{b}}\sqrt{\frac{c_{\alpha}^{2}}{\beta m_{\alpha}\omega_{\alpha}^{2}}}\bigg[ \frac{\omega_{\alpha}}{\omega_{\alpha n}}\lambda_{\alpha n}\cos{\omega_{\alpha}t} \\ 
    + \left(\xi_{\alpha n} + i\frac{\omega_{n}}{\omega_{\alpha n}}\lambda_{\alpha \bar{n}} \right) \sin{\omega_{\alpha}t} \bigg]
\end{multline}
where $\lambda_{\alpha n}$ and $\xi_{\alpha n}$ are Gaussian random variates with zero mean and unit variance and $\omega_{\alpha n}^{2}=\omega_{\alpha}^{2}+\omega_{n}^{2}$. The memory kernels of $R_{n}^{\mathbb{C}}(t)$ are
\begin{align}\nonumber
    \langle R_{n}^{\mathbb{C}}(s)R_{n}^{\mathbb{C}}(t)\rangle &= \frac{\zeta(t-s)}{\beta}-\frac{K_{n}(t-s)}{\beta}\\
\nonumber
    \langle R_{n}^{\mathbb{C}}(s)R_{\bar{n}}^{\mathbb{C}}(t)\rangle &= -\frac{iL_{n}(t-s)}{\beta}
\\
    \langle R_{n}^{\mathbb{C}}(s)R_{n^{\prime}}^{\mathbb{C}}(t)\rangle &= 0,\quad\quad |n|\ne|n'|
\end{align}
where
\begin{subequations}\label{eq:L_from_J}
\begin{align}
    K_{n}(t)     &= \frac{2\omega_{n}^{2}}{\pi}\int_{0}^{\infty}\!\diff \omega\,\frac{J(\omega)}{\omega}\frac{\cos\omega t}{\omega^{2}+\omega_{n}^{2}} \label{eq:K_from_J}\\
    L_{n}(t)     &= \frac{2\omega_{n}}{\pi}\int_{0}^{\infty}\!\diff \omega\,J(\omega)\frac{\sin\omega t}{\omega^{2}+\omega_{n}^{2}}.
\end{align}
\end{subequations}

\Eqn{eq:GLE_matsubara} differs from the GLE of Ref.~\onlinecite{prada2023comparison} only in the form of the transient driving terms in the second line.
These terms result from locally equilibrating the bath modes around the system coordinate at time $t=0$ and cannot be removed by adjusting the initial distribution of the bath modes. They arise because the non-Markovian $R_{n}^{\mathbb{C}}(t)$ lacks history at $t=0$. As time advances, the noise gradually builds up a history, and these terms consequently decay to zero. They therefore have no effect on the ability of \eqn{eq:GLE_matsubara} to equilibrate the system in the long time limit, and in what follows shall be neglected.

\section{Equilibration properties of the Matsubara GLE}\label{sec:exact_quatnum_thermostat}

This Section focuses on the main goal of the article, which is to show that the GLE of \eqn{eq:GLE_matsubara} is capable of equilibrating an arbitrary initial system phase-space distribution $\rho_0({\bf P},{\bf Q})$. We first consider the special case of a white-noise spectral density in Sec.~\ref{sec:matsubara_white_noise}, then extend to the Debye--Drude spectral density in Sec.~\ref{sec:more_general_spectral_densities_mats}. As mentioned above, the latter is sufficiently general to imply that the GLE will also equilibrate any other spectral density that can be expanded as a sum of Lorentzians. 

\begin{figure*}[t]
    \centering
    \resizebox{\textwidth}{!}{
        \input{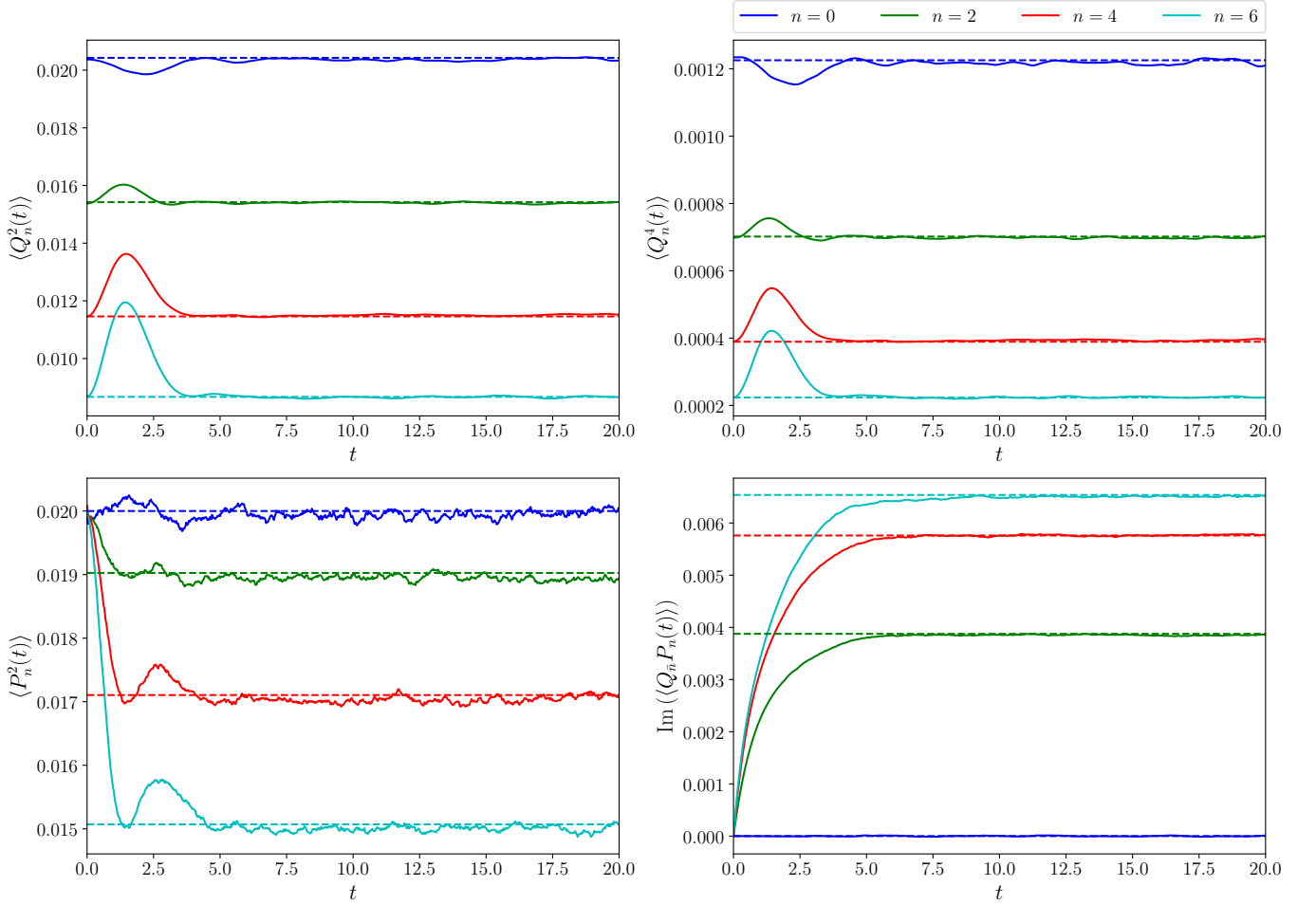}
    }
    \caption{The moments $\langle Q_{n}^{2}(t)\rangle$, $\langle Q_{n}^{4}(t)\rangle$, $\langle P_{n}^{2}(t)\rangle$ and the imaginary component of $\langle Q_{\bar{n}}P_{n}(t)\rangle$ for a quartic oscillator coupled to a white bath following the dynamics of \eqn{eq:LE_matsubara} from the initial distribution of Eq.~(\ref{eq:mats_rp_dist_white}) are plotted with solid lines. In the long-time limit, these match the corresponding moments of the quantum Boltzmann distribution of \eqn{eq:redeq} which are plotted with dotted lines.}
    \label{fig:complex_noise_white}
\end{figure*}

\subsection{White-noise spectral density}\label{sec:matsubara_white_noise}

For a white-noise spectral density, neglecting the transient terms, \eqn{eq:GLE_matsubara} simplifies to
\begin{align}\label{eq:LE_matsubara}
    m\ddot{Q}_{n}(t) &= f^{(n)}_{M}(\mathbf{Q}(t)) - m\gamma\dot{Q}_{n}(s) + R_{n}^{\mathbb{C}}(t)
\end{align}
and the non-zero components of the noise kernel become
\begin{align}\label{eq:mats_fd_relations_white}
    \langle R_{n}^{\mathbb{C}}(s)R_{n}^{\mathbb{C}}(t)\rangle &= \frac{2m\gamma}{\beta}\,\delta(t-s)-\frac{m\gamma|\omega_{n}|}{\beta}\e^{-|\omega_{n}(t-s)|}\nonumber\\
    \langle R_{n}^{\mathbb{C}}(s)R_{\bar{n}}^{\mathbb{C}}(t)\rangle &= -\frac{im\gamma\omega_{n}}{\beta}\sgn(t-s)\e^{-|\omega_{n}(t-s)|}.
\end{align}
Superficially \eqn{eq:LE_matsubara} resembles a Langevin equation. However, the dynamics it describes is non-Markovian owing to the correlations in the noise kernel. Nevertheless, by introducing a collection of auxiliary variables each undergoing an Ornstein--Uhlenbeck process to treat the noise, the dynamics can be mapped onto a Markovian stochastic process.

\subsubsection{Markovian auxiliary space and Fokker--Planck equation}

We introduce the auxiliary variables by expanding $R_{n}^{\mathbb{C}}(t)$ as 
 \begin{equation}\label{eq:white_undetermined_noise}
    R_{n}^{\mathbb{C}}(t) = a_{n}W_{n}(t) + b_{n}Z_{n}(t) + c_{n}W_{\bar{n}}(t) + d_{n}Z_{\bar{n}}(t)
\end{equation}
where the $W_{n}(t)$ are independent white noise variates with the same kernels as \eqn{eq:kernel_white}, and the $Z_n(t)$ are solutions to the Ornstein--Uhlenbeck equation
\begin{equation}\label{eq:Z_eom}
    \dot{Z}_{n}(t)=-|\omega_{n}|Z_{n}(t)+|\omega_{n}|W_{n}(t).
\end{equation}
The complex-valued coefficients $\left\{a_n,b_n,c_n,d_n \right\}$ satisfy a set of constraints. First, since $R_{n}^{\mathbb{C}}$, $W_n$ and $Z_n$ are  Matsubara variables they must satisfy the imaginary-time translation relations of \eqn{eq:mats_mode_symmetry_1}, which requires that
\begin{equation}\label{eq:noise_coeffs_negative_relations}
    a_{n}=a_{\bar{n}},\quad b_{n}=b_{\bar{n}},\quad c_{n}=-c_{\bar{n}},\quad d_{n}=-d_{\bar{n}}.
\end{equation}
Second, the kernels of \eqn{eq:white_undetermined_noise} \footnote{Either $Z_{n}(0)$ must be sampled from a Gaussian distribution with zero mean and $m\gamma|\omega_{n}|/\beta$ variance or the noise must be allowed to build up its history for the noise kernels to be as shown.},
\begin{align}
    \langle R_{n}^{\mathbb{C}}(s)R_{n}^{\mathbb{C}}(t)\rangle &= \frac{2m\gamma}{\beta}\,\delta(t-s)\left[a_{n}^{2}+c_{n}^{2}\right] \nonumber\\
    &\!\!\!\!\!\!\!\!\!\!\!\!\!\!\!\!\!\!\!\!\!\!\! +\frac{m\gamma|\omega_{n}|}{\beta}\e^{|\omega_{n}(t-s)|}\left[b_{n}^{2} + d_{n}^{2} + 2\left(a_{n}b_{n}+c_{n}d_{n}\right)\right]\nonumber\\
    \langle R_{n}^{\mathbb{C}}(s)R_{\bar{n}}^{\mathbb{C}}(t)\rangle &= - \frac{m\gamma|\omega_{n}|}{\beta}\sgn(t-s)\e^{|\omega_{n}(t-s)|} \nonumber\\
    &\quad\quad\quad\quad\times\left[2\left(a_{n}d_{n}-b_{n}c_{n}\right)\right]
\end{align}
must match those of Eq.~(\ref{eq:mats_fd_relations_white}), which requires that
\begin{align}\label{eq:noise_coeffs_relations_white}
    a_{n}^{2}+c_{n}^{2}&=1\nonumber\\
    b_{n}^{2} + d_{n}^{2} + 2\left(a_{n}b_{n}+c_{n}d_{n}\right) &= -1\nonumber\\
    a_{n}d_{n}-b_{n}c_{n} &= \frac{i\sgn(n)}{2}.
\end{align}
These constraints, together with \eqn{eq:noise_coeffs_negative_relations}, are not sufficient to specify the coefficients uniquely, so we have some freedom in how they are chosen. A convenient choice, used to obtain the numerical results in Sec.~\ref{sec:simple_numerical_test}, is \begin{align}
a_n = 1,\quad b_n =-\frac{3}{2},\quad c_n = 0,\quad d_n = \frac{i\sgn(n)}{2}.
\end{align}

Using \eqn{eq:white_undetermined_noise}, we can now rewrite the dynamics of \eqn{eq:LE_matsubara} as
\begin{align}\label{eq:markovian_LE}
    \dot{Q}_{n}(t) &= \frac{P_{n}(t)}{m}\nonumber\\
    \dot{P}_{n}(t) &= -\frac{\partial U_{M}(\mathbf{Q}(t))}{\partial Q_{n}}-\gamma P_{n}(t) + \chi_{n}(t) + a_{n}W_{n}(t) \nonumber\\
    &\quad\quad\quad\quad\quad+ c_{n}W_{\bar{n}}(t)\nonumber\\
    \dot{\chi}_{n}(t) &= -|\omega_{n}|\chi_{n}(t) + |\omega_{n}|\left[b_{n}W_{n}(t)+d_{n}W_{\bar{n}}(t)\right]
\end{align}
where we have defined
\begin{equation}\label{eq:combining_Zs}
    \chi_{n}(t) = b_{n}Z_{n}(t)+d_{n}Z_{\bar{n}}(t).
\end{equation}
These equations of motion are Markovian and of the form of \eqn{eq:general_eom_continuous}. We can therefore  construct the matrix  
\begin{equation}\label{eq:white_noise_matrix}
    \mathbf{\Theta} = \frac{m\gamma}{\beta}\left(\begin{matrix}
        \mathbf{I} && \mathbf{0} &&\boldsymbol{\kappa} \\
        \mathbf{0} && \mathbf{0} && \mathbf{0} \\
         \boldsymbol{\kappa}^{T} && \mathbf{0} && \boldsymbol{\iota}
    \end{matrix}\right)
\end{equation}
indexed by $({\bf P},{\bf Q} ,{\boldsymbol{\chi}})$, in which the $M$-dimensional submatrix elements are
\begin{align}
{{\iota}}_{nn'} &= \left(b_{n}^{2}+d_{n}^{2}\right)\omega_{n}^{2} \delta_{nn'}\nonumber\\
{{\kappa}}_{nn'}  &=(a_{n}b_{n}+c_{n}d_{n})|\omega_{n}|\delta_{nn'}-i{\omega_{n}\over2}\delta_{\bar{n} n'}
\end{align}
from which we obtain the Fokker--Planck equation
\begin{multline}\label{eq:fp_mats_white}
    \frac{\partial \rho(\mathbf{P}, \mathbf{Q}, \boldsymbol{\chi}, t)}{\partial t} = \sum_{n=-\widetilde{M}}^{\widetilde{M}}\bigg[-\frac{\partial}{\partial Q_{n}}\frac{P_{n}}{m} \\
    + \frac{\partial}{\partial P_{n}}\left(\frac{\partial U_{M}(\mathbf{Q})}{\partial Q_{n}} - \chi_{n}\right) + \gamma \frac{\partial}{\partial P_{n}}\left(P_{n}+\frac{m}{\beta}\frac{\partial}{\partial P_{n}}\right) \\
    + |\omega_{n}| \frac{\partial}{\partial \chi_{n}}\left(\chi_{n}+\frac{m\gamma|\omega_{n}|}{\beta}\left(b_{n}^{2}+d_{n}^{2}\right)\frac{\partial}{\partial \chi_{n}}\right) \\
    + \frac{2m\gamma|\omega_{n}|}{\beta}\left(a_{n}b_{n}+c_{n}d_{n}\right) \frac{\partial^{2}}{\partial P_{n}\partial \chi_{n}} \\- \frac{i m \gamma\omega_{n}}{\beta}\frac{\partial^{2}}{\partial P_{n}\partial\chi_{\bar{n}}} \bigg]\rho(\mathbf{P}, \mathbf{Q}, \boldsymbol{\chi}, t).
\end{multline}

\subsubsection{Form of the FPE stationary solution}\label{sec:matsubara_stationary_distribution_white}

To demonstrate that \eqn{eq:LE_matsubara} will equilibrate any initial distribution $\rho_0({\bf P},{\bf Q})$ given sufficient time, we need to find the stationary solution of the FPE in \eqn{eq:fp_mats_white} and show that it marginalises to $\rho_\text{eq}({\bf P},{\bf Q})$ of  \eqn{eq:redeq}. 

Finding stationary solutions to FPEs is in general difficult \cite{risken1989fokker}. However, \eqn{eq:fp_mats_white} turns out to be readily solvable because one can first compute the solution in the special case that $V(q)$ is harmonic. This is easily done using the method of moments \cite{risken1989fokker}, as described in Ref.~\onlinecite{moore2026matsubara}. One then finds that the harmonic $V(q)$ reappears in the stationary solution and replacing it by the general case \footnote{By which we mean that $V(q)$ is unspecified, anharmonic, and bound.} gives
\begin{equation}\label{eq:mats_dist_harmonic_white_full}
    \rho_\text{stat}(\mathbf{P},\mathbf{Q},\boldsymbol{\chi}) ={1\over \cal{N}} \e^{-\beta\Phi(\mathbf{P},\mathbf{Q},\boldsymbol{\chi})} \delta\left(\chi_{0}\right)
\end{equation}
in which
\begin{multline}\label{eq:mats_dist_exponent_general_white_full}
    \Phi(\mathbf{P},\mathbf{Q},\boldsymbol{\chi}) = F_{M}(\mathbf{P},\mathbf{Q}) -i\theta_{M}(\mathbf{P},\mathbf{Q}) \\ 
      +\sum_{n=-\widetilde{M}}^{\widetilde{M}}\!\!\!\!\!\!\!\!{\phantom{\frac{x}{y}}}^{\prime}\,\,\,  {1\over b_{n}^{2}+d_{n}^{2}}\left[\frac{\chi_{n}^{2}}{2m\gamma|\omega_{n}|}
    + \chi_{n}\left(Q_{n}-i\,\text{sgn}(n)Q_{\bar{n}}\right) \right]
\end{multline}
where $F_{M}(\mathbf{P},\mathbf{Q})$ is defined in \eqn{eq:fm} and the prime indicates that $n=0$ is omitted from the sum. Substituting into \eqn{eq:fp_mats_white}, we obtain
\begin{align}\label{eq:drho_dt}
    \frac{\partial \rho_\text{stat}}{\partial t} &= i\beta\rho_\text{stat}\sum_{n=-\widetilde{M}}^{\widetilde{M}}\omega_{n}Q_{\bar{n}}\frac{\partial U_{M}}{\partial Q_{n}}\nonumber\\
    &=0
\end{align}
where the second equality follows from the imaginary-time translation symmetry of $U_{M}$---see \eqn{vint}. Thus $\rho_{\mathrm{stat}}$ is the general form of the stationary solution to the FPE of \eqn{eq:fp_mats_white}. On marginalising over $\boldsymbol{\chi}$, we obtain
\begin{equation}
     \int\!d\boldsymbol{\chi}\, \rho_\text{stat}(\mathbf{P},\mathbf{Q},\boldsymbol{\chi})=\rho_\text{eq}({\bf P},{\bf Q}).
\end{equation}

We have therefore shown that the white-noise Matsubara GLE of \eqn{eq:LE_matsubara} is capable of equilibrating an arbitrary initial distribution of system phase-space points to the distribution $\rho_\text{eq}({\bf P},{\bf Q})$ of \eqn{eq:redeq}. 

\subsubsection{The imaginary momentum--position correlation}

The Matsubara Langevin equation of \eqn{eq:LE_matsubara} produces the equilibrium distribution
 $\rho_\text{eq}({\bf P},{\bf Q})$ without explicitly invoking the phase $\exp[-i\theta_M({\bf P},{\bf Q})]$. In other words, an 
 expectation value of some property $A({\bf P},{\bf Q})$ can be evaluated using 
 \begin{multline}\label{eq:langap}
\int\! d{\bf P}\int\! d{\bf Q}\, \rho_\text{eq}({\bf P},{\bf Q}) A({\bf P},{\bf Q})=\\\lim_{N_\text{t}\to\infty}{1\over N_\text{t}}\sum_{i=1}^{N_\text{t}} A({\bf P}(t_i),{\bf Q}(t_i))
\end{multline}
where the phase-space points $\left\{{\bf P}(t_i),{\bf Q}(t_i) \right\}$ are taken at $N_\text{t}$ times $t_i$ from a trajectory propagated using \eqn{eq:LE_matsubara}. The phase is included naturally in the stationary distribution as a result of the complex-valued random force $R_{n}^{\mathbb{C}}$ which pushes the phase-space variables away from the real axis, such that $P_n$ and $Q_{\bar{n}}$ maintain the purely imaginary correlation given by $\theta_M({\bf P},{\bf Q})$. 

Another way of looking at this is that the propagation of \eqn{eq:LE_matsubara} has the effect of analytically continuing the distribution $\rho_\text{eq}({\bf P},{\bf Q})$ into the complex plane, such that the phase no longer appears. It therefore eliminates the dreaded `phase problem' but has replaced it with another numerical difficulty, since classical dynamics in the complex plane is well known to be numerically unstable. From \eqn{eq:markovian_LE}, we can see that the instability is likely to be tamed by increasing $\gamma$ and worsened by increasing $|n|$. This suggests that any attempt to propagate trajectories using \eqn{eq:LE_matsubara} will be numerically stable only if $M$ is sufficiently small.

\subsubsection{A simple numerical test}\label{sec:simple_numerical_test}

We have investigated these numerical properties for the simple case of a quartic system potential
\begin{equation}\label{eq:quartic_potential}
    V(q)=\frac{q^{4}}{4}
\end{equation}
with a moderate damping strength of $\gamma=1$. At a temperature of $\beta=50$, we find that the dynamics is stable for $M=13$, which is far lower than the value of 
$M\simeq 100$ needed to converge $\rho_\text{eq}({\bf P},{\bf Q})$ at this temperature. Nonetheless, this small value of $M$ is  sufficient to confirm numerically that \eqn{eq:LE_matsubara} is capable of generating the imaginary momentum--position correlations in $\rho_\text{eq}(\mathbf{P},\mathbf{Q})$. A total of $2\times 10^{5}$ phase-space points were sampled from an initial distribution
  \begin{multline}\label{eq:mats_rp_dist_white}
    \rho_{0}(\mathbf{P},\mathbf{Q}) ={1\over\cal{N}} \exp\Bigg(-\beta\Bigg[ U_{M}(\mathbf{Q}) + \sum_{n=-\widetilde{M}}^{\widetilde{M}} \bigg[ \frac{P_{n}^{2}}{2m} \\
    + \frac{1}{2}m\left(\omega_{n}^{2} + \gamma|\omega_{n}|\right)Q_{n}^{2}\bigg]\Bigg]\Bigg)
\end{multline}
where $m=1$ and $\hbar=1$. The points were propagated until $t=20$ (reduced units) by integrating \eqn{eq:markovian_LE} using a modified velocity Verlet algorithm with time-step $\Delta t= 0.005$. Figure~1 plots various moments of $\rho(\mathbf{Q},\mathbf{P},t)$  against time, with $\langle P_nQ_{\bar{n}} \rangle$ illustrating clearly the growth of the imaginary momentum--position correlation between $P_n$ and $Q_{\bar{n}}$.

\subsection{Debye--Drude spectral density}\label{sec:more_general_spectral_densities_mats}

For the Debye--Drude spectral density, the derivation follows the same steps as those just given for white noise, with the addition of extra auxiliary variables to account for the non-Markovian decay of $\zeta(t)$.

The Debye--Drude noise kernels, obtained by substituting $J(\omega)$ of \eqn{eq:general_spectral_density} into \eqn{eq:L_from_J}, are
\begin{align}\label{eq:mats_fd_relations_nn_general}
    \langle R_{n}^{\mathbb{C}}(s)R_{n}^{\mathbb{C}}(t)\rangle &= \frac{m}{\beta}\frac{\gamma\omega_{c}^{2}}{\omega_{c}^{2}-\omega_{n}^{2}}\nonumber\\
    &\quad\quad\times\left(\omega_{c}\e^{-\omega_{c}|t-s|}-|\omega_{n}|\e^{-|\omega_{n}(t-s)|}\right)\nonumber\\
    \langle R_{n}^{\mathbb{C}}(s)R_{\bar{n}}^{\mathbb{C}}(t)\rangle &= \frac{im\omega_{n}}{\beta}\sgn(t-s)\frac{\gamma\omega_{c}^{2}}{\omega_{c}^{2}-\omega_{n}^{2}}\nonumber\\
    &\quad\quad\times\left(\e^{-\omega_{c}|t-s|}-\e^{-|\omega_{n}(t-s)|}\right) 
\end{align}
and can be reproduced by expanding the noise as
\begin{multline}\label{eq:general_undetermined_noise}
    R_{n}^{\mathbb{C}}(t) = a_{n}Y_{n}(t) + b_{n}Z_{n}(t) + c_{n}Y_{\bar{n}}(t) + d_{n}Z_{\bar{n}}(t)
\end{multline}
where the $Z_n(t)$ are the solutions of \eqn{eq:Z_eom}, and the $Y_n(t)$ (analogous to $y(t)$ of Sec.~\ref{sec:classical_auxiliary_variables}) are the solutions of
\begin{equation}\label{eq:Yi_eom}
    \dot{Y}_{n}(t)=-\omega_{c}Y_{n}(t)+\omega_{c}W_{n}(t).
\end{equation}
The coefficients $\left\{a_n,b_n,c_n,d_n \right\}$ satisfy the  symmetry constraints given in \eqn{eq:noise_coeffs_negative_relations}. Additional constraints are found by matching the kernels to \eqn{eq:mats_fd_relations_nn_general} to obtain \footnote{For the noise in \eqn{eq:general_undetermined_noise} to have the desired kernels, the auxiliary variables must also be sampled with initial conditions $\langle Y_{j,n}(0)Y_{j^{\prime},n^{\prime}}(0)\rangle=m\gamma_{j}\omega_{c,j}\delta_{nn^{\prime}}\delta_{jj^{\prime}}/\beta$, $\langle Z_{j,n}(0)Z_{j^{\prime},n^{\prime}}(0)\rangle=m\gamma_{j}|\omega_{n}|\delta_{nn^{\prime}}\delta_{jj^{\prime}}/\beta$ and $\langle Y_{j,n}(0)Z_{j^{\prime},n^{\prime}}(0)\rangle = 2m\gamma_{j}\omega_{c,j}|\omega_{n}|\delta_{nn^{\prime}}\delta_{jj^{\prime}}/\beta(\omega_{c,j}+|\omega_{n}|)$. Alternatively, the noise can simply be allowed to build up its history.}
\begin{align}
    a_{n}^{2}+c_{n}^{2} + \frac{2|\omega_{n}|}{\omega_{c}+|\omega_{n}|}\left(a_{n}b_{n} + c_{n}d_{n}\right)&=\frac{\omega_{c}^{2}}{\omega_{c}^{2}-\omega_{n}^{2}}\nonumber\\
    b_{n}^{2} + d_{n}^{2} + \frac{2\omega_{c}}{\omega_{c}+|\omega_{n}|}\left(a_{n}b_{n}+c_{n}d_{n}\right) &= -\frac{\omega_{c}^{2}}{\omega_{c}^{2}-\omega_{n}^{2}}\nonumber\\
    a_{n}d_{n}-b_{n}c_{n} = \frac{i\,\omega_{c}\sgn(n)}{2\left(\omega_{c}-|\omega_{n}|\right)}&.
\end{align}
To deal with the non-Markovian friction term in the GLE, we introduce the variables $V_{n}(t)$ (analogous to $v(t)$ of Sec.~\ref{sec:classical_auxiliary_variables}) which satisfy
\begin{equation}
    \dot{V}_{n}(t)=-\omega_{c}V_{n}(t) + m\gamma \omega_{c} \dot{Q}_{n}(t)
\end{equation}
with initial conditions $V_{n}(0)=0$. As a result,
\begin{equation}
    V_{n}(t) = m\gamma\omega_{c}\int_{0}^{t}\!\diff s\,\e^{-\omega_{c}(t-s)}\dot{Q}_{n}(s)
\end{equation}
which regenerates the corresponding friction term in \eqn{eq:GLE_matsubara} (for the Debye--Drude spectral density). Taking the linear combination
\begin{equation}\label{eq:combining_V_Ys}
    \psi_{n}(t)=-V_{n}(t)+a_{n}Y_{n}(t)+c_{n}Y_{\bar{n}}(t)
\end{equation}
\begin{widetext}
and defining $\chi_{n}$ as in \eqn{eq:combining_Zs}, we can then rewrite the stochastic dynamics as 
\begin{align}
    \dot{Q}_{n}(t) &= \frac{P_{n}(t)}{m}\nonumber\\
    \dot{P}_{n}(t) &= -\frac{\partial U_{M}(\mathbf{Q}(t))}{\partial Q_{n}} +\psi_{n}(t)+\chi_{n}(t)\nonumber\\
    \dot{\psi}_{n}(t) &= -\omega_{c}\psi_{n}(t) - \gamma\omega_{c}P_{n}(t) 
    + \omega_{c}\left[a_{n}W_{n}(t) +  c_{n}W_{\bar{n}}(t)\right]\nonumber\\
    \dot{\chi}_{n}(t) &= -|\omega_{n}|\chi_{n}(t) + |\omega_{n}|\left[b_{n}W_{n}(t)+d_{n}W_{\bar{n}}(t)\right]
\end{align}
which is now Markovian in the extended space $(\mathbf{P}, \mathbf{Q}, \boldsymbol{\psi}, \boldsymbol{\chi})$.
Constructing the matrix $\boldsymbol{\Theta}$, we obtain the Debye--Drude Matusbara FPE,
\begin{multline}\label{eq:fp_mats_general}
    \frac{\partial \rho(\mathbf{P}, \mathbf{Q}, \boldsymbol{\psi}, \boldsymbol{\chi}, t)}{\partial t} = \sum_{n=-\widetilde{M}}^{\widetilde{M}}\Bigg[-\frac{\partial}{\partial Q_{n}}\frac{P_{n}}{m} + \frac{\partial}{\partial P_{n}}\left(\frac{\partial U_{M}(\mathbf{Q})}{\partial Q_{n}} -\psi_{n}-\chi_{n}\right) + \gamma\omega_{c}\frac{\partial}{\partial \psi_{n}}P_{n} \\
    + \omega_{c}\frac{\partial}{\partial \psi_{n}}\left(\psi_{n}+\frac{m\gamma\omega_{c}}{\beta}\left(a_{n}^{2}+c_{n}^{2}\right)\frac{\partial}{\partial\psi_{n}}\right) + |\omega_{n}| \frac{\partial}{\partial \chi_{n}}\left(\chi_{n}+\frac{m\gamma|\omega_{n}|}{\beta}\left(b_{n}^{2}+d_{n}^{2}\right)\frac{\partial}{\partial \chi_{n}}\right) \\
    + \frac{2m\gamma\omega_{c}|\omega_{n}|}{\beta}\left(a_{n}b_{n}+c_{n}d_{n}\right)\frac{\partial^{2}}{\partial \psi_{n}\partial \chi_{n}} - \frac{i m\gamma\omega_{c}^{2}\omega_{n}}{\beta\left(\omega_{c}-|\omega_{n}|\right)}\frac{\partial^{2}}{\partial \psi_{n}\partial \chi_{\bar{n}}}\Bigg]\rho(\mathbf{P}, \mathbf{Q}, \boldsymbol{\psi}, \boldsymbol{\chi}, t).
\end{multline}
The stationary state $\rho_{\mathrm{stat}}$ can be constructed by following the same steps as for the white bath (i.e.~first solving for a hamonic potential using the method of moments \cite{moore2026matsubara}, then replacing the harmonic potential by a general anharmonic $V(q)$), and is found to be
\begin{equation}\label{eq:general_matsubara_dist}
    \rho_{\mathrm{stat}}(\mathbf{P},\mathbf{Q},\boldsymbol{\phi}, \boldsymbol{\chi}) ={1\over\cal{N}} \e^{-\beta\Phi(\mathbf{Q},\mathbf{P},\boldsymbol{\phi}, \boldsymbol{\chi})} \delta\left(\chi_{0}\right)
\end{equation}
where, 
\begin{multline}\label{eq:general_general_exponent}
    \Phi(\mathbf{P}, \mathbf{Q}, \boldsymbol{\phi}, \boldsymbol{\chi}) = H_M({\bf P},{\bf Q}) -i\theta_{M}(\mathbf{P}, \mathbf{Q}) + \sum_{n=-\tilde{M}}^{\tilde{M}}  \left[\frac{1}{2}m|\omega_{n}|\gamma Q_{n}^{2}   + \frac{|\omega_{n}|}{\omega_{c}}Q_{n}\psi_{n}+ \frac{\omega_{c}+|\omega_{n}|}{2m\gamma\omega_{c}^{2}}\psi_{n}^{2}\right] \\
    +\sum_{n=-\tilde{M}}^{\tilde{M}}\!\!\!\!\!\!\!\!\!{\phantom{\frac{x}{y}}}^{\prime}\,\left\{{a_{n}^{2}+c_{n}^{2}\over b_{n}^{2}+d_{n}^{2}}\left[\frac{\omega_{c}+|\omega_{n}|}{2 m\gamma\omega_{c}|\omega_{n}|}\chi_{n}^{2}
    + \left(Q_{n} + \frac{\psi_{n}}{ m\gamma\omega_{c}}\right)\left(\chi_{n} + \frac{i\omega_{c}\sgn(n)}{\left(\omega_{c}-|\omega_{n}|\right)(a_{n}^{2}+c_{n}^{2})}\chi_{\bar{n}}\right)\right]+ \frac{\psi_{n}\chi_{n}}{m\gamma\omega_{c}}\right\}.
\end{multline}
\end{widetext}
Substituting $\rho_{\mathrm{stat}}$ into \eqn{eq:fp_mats_general} then yields the expression given on the right-hand side of \eqn{eq:drho_dt}, thus confirming that $\rho_{\mathrm{stat}}$ is the stationary solution. Finally, marginalising
 over $(\boldsymbol{\psi}, \boldsymbol{\chi})$ gives $\rho_\text{eq}({\bf P},{\bf Q})$ of 
Eq.~(\ref{eq:redeq}), with
\begin{equation}
    \tilde{\zeta}(|\omega_{n}|) = m\frac{ \gamma\omega_{c}}{\omega_{c} + |\omega_{n}|}
\end{equation}
which is the expression for $\tilde{\zeta}(|\omega_{n}|)$ obtained by substituting the Debye--Drude spectral density [\eqn{eq:general_spectral_density}] into \eqn{eq:zeta_hat}.

We have thus shown that the Matsubara GLE of \eqn{eq:GLE_matsubara} equilibrates to the thermal equlibrium state $\rho_\text{eq}({\bf P},{\bf Q})$  for a Debye--Drude spectral density. As mentioned above, this is sufficiently general to cover any spectral density that can be expanded as a sum of Lorentzians.

\section{Conclusions}\label{sec:conc}

It is surprising that stochastic classical trajectories in an extended space can equilibrate to the exact quantum equilibrium state, correctly including the purely imaginary correlations between the momentum and position variables. Nonetheless, there are no free lunches, and the stochastic classical dynamics produces these correlations by evolving the trajectories into the complex plane.  Inevitably, this leads to numerical instability, so we are not advocating the use of this approach as a practical method. It may, however, be useful as a starting point from which to derive more approximate and practical methods for simulating continuous-variable open quantum dynamics.

\begin{acknowledgments}
   W.H.D.M.\ acknowledges the UK Engineering and Physical Sciences Research Council for supporting this work, and is grateful for grants from the Cambridge Philosophical Society and Peterhouse, Cambridge.
\end{acknowledgments}

\section*{Data availability}
   The data that support the findings of this article are publicly available in Ref.~\onlinecite{moore2026research}.

\appendix
\section{Imaginary-time symmetries}\label{app:immaginary_time_symmetries}
The effect of an imaginary-time translation on a variable $r(\tau)$ (where $r$ is $p$ or $q$ and $R_{n}\equiv R_{n}\left(\tau_{0}=0\right)$) is 
\begin{multline}
    r(\tau_{0}+\tau) = R_{0}+\sqrt{2}\sum_{n=1}^{\tilde{M}}\big[R_{n}\sin\left(\omega_{n}\left(\tau_{0}+\tau\right)\right) \\
    + R_{\bar{n}}\cos\left(\omega_{n}\left(\tau_{0}+\tau\right)\right)\big]
\end{multline}
which can be rearranged into
 \begin{multline}
    r(\tau_{0}+\tau)
    = R_{0}(\tau_{0})+\sqrt{2}\sum_{n=1}^{\tilde{M}}\big[R_{n}(\tau_{0})\sin\omega_{n}\tau \\
    + R_{\bar{n}}(\tau_{0})\cos\omega_{n}\tau\big]
\end{multline}
where
\begin{subequations}
\begin{align}
    R_{n}(\tau_{0}) &= R_{n}\cos\omega_{n}\tau_{0} -  R_{\bar{n}}\sin\omega_{n}\tau_{0}\label{eq:Qn_imag_time}\\
    R_{\bar{n}}(\tau_{0}) &= R_{n}\sin\omega_{n}\tau_{0} +  R_{\bar{n}}\cos\omega_{n}\tau_{0}.
\end{align}
\end{subequations}
Since $\omega_{n}=2n\pi/\beta\hbar$, it follows that
\begin{align}
    R_{n}(\tau_{0} + \beta\hbar/4n) &= -R_{n}\sin\omega_{n}\tau_{0} - R_{\bar{n}}\cos\omega_{n}\tau_{0} \nonumber\\
    R_{\bar{n}}(\tau_{0} + \beta\hbar/4n) &= R_{n}\cos\omega_{n}\tau_{0} -  R_{\bar{n}}\sin\omega_{n}\tau_{0} 
\end{align}
from which we obtain Eq.~(\ref{eq:mats_mode_symmetry_1}).

Equation~(\ref{eq:Qn_imag_time}) also gives
\begin{equation}\label{eq:Xn_imag_time_derivative}
    \frac{\partial R_{n}(\tau_{0})}{\partial\tau_{0}} = -\omega_{n}R_{n}\sin\omega_{n}\tau_{0} -  \omega_{n}R_{\bar{n}}\cos\omega_{n}\tau_{0}
\end{equation}
so that
\begin{equation}
  \frac{\partial R_{n}}{\partial\tau_{0}} \equiv  \frac{\partial R_{n}(0)}{\partial\tau_{0}} = - \omega_{n}R_{\bar{n}}.
\end{equation}
Subsituting this last expression (with $R\to Q$) into
\begin{equation}
{\partial U_M({\bf Q})\over\partial \tau_0}= 0 = \sum_{n=-\widetilde{M}}^{\widetilde{M}} {\partial Q_n\over\partial \tau_0}{\partial U_M({\bf Q})\over\partial Q_n}
\end{equation}
gives Eq.~(\ref{vint}).

\section{Derivation of the Matsubara GLE}\label{app:gle_derivation}

The Matsubara Hamiltonian corresponding to \eqn{eq:caldeira_leggett_classical} is
\begin{multline}
H_M^\text{tot}(\mathbf{P},\mathbf{Q},{\bf P}_\text{b},{\bf X})=H_M(\mathbf{P},\mathbf{Q}) + \sum_{\alpha=1}^{n_{b}}\sum_{n=-\widetilde{M}}^{\widetilde{M}}\Bigg[ \frac{P_{\alpha,n}^{2}}{2m_{\alpha}} \\
    + \frac{1}{2} m_{\alpha} \omega_{\alpha}^{2} \left( X_{\alpha,n}  - \frac{c_{\alpha}}{m_{\alpha}\omega_{\alpha}^{2}} Q_n \right)^{2}\Bigg]
\end{multline}
where $(P_{\alpha,n} ,X_{\alpha,n})$ are the Matsubara modes corresponding to bath variables $(p_\alpha,x_\alpha)$. To obtain an initial distribution in which the bath is locally equilibrated around the system variables, and the system is prepared in some arbitrary initial density $\rho_0$, we start with the full system--bath equilibrium distribution
\begin{equation}
   \rho_M({\bf P},{\bf Q},{\bf P}_\text{b},\mathbf{X})={1\over \cal{N}}\e^{-\beta\left[H_M^\text{tot}({\bf P},{\bf Q},{\bf P}_\text{b},\mathbf{X}) -i\theta^\text{tot}_{M}(\mathbf{P}, \mathbf{Q},{\bf P}_\text{b},\mathbf{X}) \right]}
\end{equation}
where
\begin{multline}
\theta^\text{tot}_{M}(\mathbf{P}, \mathbf{Q},{\bf P}_\text{b},\mathbf{X})= \theta_{M}(\mathbf{P}, \mathbf{Q}) \\+ \sum_{\alpha=1}^{n_{b}}\sum_{n=-\widetilde{M}}^{\widetilde{M}}\omega_{n}X_{\alpha,{\bar{n}}}P_{\alpha,n}.
\end{multline}
We then make the substitution
\begin{equation}
    P_{\alpha,n}=\Pi_{\alpha, n} + im_{\alpha}\omega_{n}X_{\alpha \bar{n}}
\end{equation}
which is equivalent to shifting the momentum integration contour so as to eliminate the bath contribution to the phase, giving
\begin{equation}\label{eq:equil_dist_bath_AC}
   \rho_M({\bf P},{\bf Q},\mathbf{\Pi}_{b},\mathbf{X})={1\over \cal{N}}\e^{-\beta\left[F_M({\bf P},{\bf Q})+ B_M({\bf \Pi}_{b},{\bf X};{\bf Q}) -i\theta_{M}(\mathbf{P}, \mathbf{Q}) \right]}
\end{equation}
where
\begin{equation}
    B_M({\bf \Pi}_{b},{\bf X};{\bf Q}) = \sum_{\alpha=1}^{n_{b}}\sum_{n=-\widetilde{M}}^{\widetilde{M}} \left(\frac{\Pi_{\alpha,n}^{2}}{2m_{\alpha}}
    + \frac{1}{2}m_{\alpha}\omega_{\alpha n}^{2}{\overline X}_{\alpha,n}^2 \right)
\end{equation}
and
\begin{equation}
{\overline X}_{\alpha,n}\equiv X_{\alpha,n} - \frac{c_{\alpha}}{m_{\alpha}\omega_{\alpha n}^{2}}Q_{n}.
\end{equation}
Replacing the system-dependent part of \eqn{eq:equil_dist_bath_AC} by $\rho_0({\bf P},{\bf Q})$, we then obtain
\begin{equation}\label{eq:app_qb_dist_bath_AC}
   \rho_M({\bf P},{\bf Q},\mathbf{\Pi}_{b},\mathbf{X})={\rho_0({\bf P},{\bf Q})\over \cal{N}}\e^{-\beta B_M({\bf \Pi}_{b},{\bf X};{\bf Q})}
\end{equation}
which describes an initial distribution in which the bath variables $X_{\alpha,n}$ are equilibrated about the system coordinate $Q_n$, and the system distribution is arbitrary.

From \eqn{eq:matty},  $H_M^\text{tot}(\mathbf{P},\mathbf{Q},{\bf P}_\text{b},{\bf X})$ generates the Matsubara-dynamics equations of motion
\begin{align}\label{eq:app_eoms}
    m\ddot{Q}_{n} &= f^{(n)}_{M}(\mathbf{Q}) + \sum_{\alpha=1}^{n_{b}}c_{\alpha}\left(X_{\alpha n} -  \frac{c_{\alpha}}{m_{\alpha}\omega_{\alpha}^{2}}Q_{n}\right)\nonumber\\
    \ddot{X}_{\alpha n} &= -\omega_{\alpha}^{2}\left(X_{\alpha n} - \frac{c_{\alpha}}{m_{\alpha}\omega_{\alpha}^{2}}Q_{n}\right)
\end{align}
with $f^{(n)}_{M}(\mathbf{Q})$ defined in \eqn{eq:force}. 
Solving for the bath dynamics up to time $t$, we obtain
\begin{multline}\label{eq:app_mats_GLE_derivation}
    m\ddot{Q}_{n}(t) = f^{(n)}_{M}(\mathbf{Q}(t)) - \int_{0}^{t}\!\diff s\, \zeta(t-s)\dot{Q}_{n}(s)
    \\ + \sum_{\alpha=1}^{n_{b}}c_{\alpha}\left[\left(X_{\alpha, n} - \frac{c_{\alpha}}{m_{\alpha}\omega_{\alpha}^{2}} Q_{n} \right)\cos{\omega_{\alpha}t} + \frac{P_{\alpha, n}}{m_{\alpha}\omega_{\alpha}}\sin{\omega_{\alpha}t}\right]
\end{multline}
with $\zeta(t)$ given in \eqn{eq:kernelfromJintegral}, $X_{\alpha,n}\equiv X_{\alpha,n}(t=0)$ and similarly for
$Q_n$ and $P_{\alpha,n}$. Converting from ${\bf P}_\text{b}$ to ${\bf \Pi}_\text{b}$, we then obtain
\begin{multline}
    m\ddot{Q}_{n}(t) = f^{(n)}_{M}(\mathbf{Q}(t)) - \int_{0}^{t}\!\diff s\, \zeta(t-s)\dot{Q}_{n}(s) \\+ \sum_{\alpha=1}^{n_{b}}c_{\alpha}\bigg[ {\overline X}_{\alpha,n} \cos{\omega_{\alpha}t} 
    + \left(\frac{\Pi_{\alpha, n}}{m_{\alpha}\omega_{\alpha}} + i \frac{\omega_{n}}{\omega_{\alpha}}{\overline X}_{\alpha, \bar{n}}\right)\sin{\omega_{\alpha}t} \\- \frac{c_{\alpha}\omega_{n}}{m_{\alpha}\omega_{\alpha}\omega_{\alpha n}^{2}}\left(\frac{\omega_{n}}{\omega_{\alpha}}Q_{n}\cos\omega_{\alpha}t - iQ_{\bar{n}}\sin\omega_{\alpha}t\right) \bigg].
\end{multline}
Comparison of the variances of ${\bf \Pi}_\text{b}$ and ${\bf X}$ in \eqn{eq:app_qb_dist_bath_AC} with \eqn{eq:complex_noise} shows that the first three terms in the sum correspond to $R_{n}^{\mathbb{C}}(t)$ in the GLE [\eqn{eq:GLE_matsubara}]; comparison with \eqn{eq:L_from_J} shows that the last two terms correspond to $- \left[Q_{n}K_{n}(t) - iQ_{\bar{n}}L_{n}(t) \right]$.

\bibliographystyle{apsrev4-2}
\bibliography{References}

@string{JCP         = "J. Chem. Phys."}

@string{RMP         = "Rev. Mod. Phys."}

@article{Caldeira1983,
  author  = {Caldeira, A. O. and Leggett, A. J.},
  journal = {Physica A},
  volume  = {121},
  pages   = {587--616},
  year    = {1983}
}

@article{TorresMiyares2022,
  author  = {Torres-Miyares, E. E. and Rojas-Lorenzo, G. and Rubayo-Soneira, J. and Miret-Art{\'e}s, S.},
  journal = {Phys. Chem. Chem. Phys.},
  volume  = {24},
  pages   = {15871--15890},
  year    = {2022}
}

@article{miret2005dynamics,
  title={The dynamics of activated surface diffusion},
  author={Miret-Art{\'e}s, Salvador and Pollak, Eli},
  journal={J. Phys.: Condens. Matter},
  volume={17},
  number={49},
  pages={S4133--S4150},
  year={2005}
}

@article{Trenins2025,
  author  = {Trenins, G. and Rossi, M.},
  journal = {Phys. Rev. Lett.},
  volume  = {134},
  pages   = {226201},
  year    = {2025}
}

@article{Tanimura1993,
  author  = {Tanimura, Y. and Mukamel, S.},
  journal = {J. Chem. Phys.},
  volume  = {99},
  pages   = {9496--9511},
  year    = {1993}
}

@article{Kato2004,
  author  = {Kato, T. and Tanimura, Y.},
  journal = {J. Chem. Phys.},
  volume  = {120},
  pages   = {260--271},
  year    = {2004}
}

@article{Grabert1988,
  author  = {Grabert, H. and Schramm, P. and Ingold, G.-L.},
  journal = {Phys. Rep.},
  volume  = {168},
  pages   = {115--207},
  year    = {1988}
}

@article{Chandler1981,
  author  = {Chandler, D. and Wolynes, P. G.},
  journal = {J. Chem. Phys.},
  volume  = {74},
  pages   = {4078--4095},
  year    = {1981}
}

@phdthesis{moore2026matsubara,
  title={Matsubara dynamics for open quantum systems},
  author={Moore, William H. D.},
  school={University of Cambridge},
  year={2026}
}

@book{tuckerman2010statistical,
  title={Statistical Mechanics: Theory and Molecular Simulation},
  author={Tuckerman, Mark},
  year={2010},
  publisher={Oxford University Press}
}

@book{weiss2012quantum,
  title={Quantum Dissipative Systems},
  author={Weiss, Ulrich},
  year={2012},
  publisher={World Scientific}
}

@article{Althorpe2021,
  author  = {Althorpe, S. C.},
  journal = {Eur. Phys. J. B},
  volume  = {94},
  pages   = {155},
  year    = {2021}
}

@article{Lawrence2019,
  title={On the calculation of quantum mechanical electron transfer rates},
  author={Lawrence, Joseph E and Fletcher, Theo and Lindoy, Lachlan P and Manolopoulos, David E},
  journal = {J. Chem. Phys.},
  volume  = {151},
  pages   = {114119},
  year    = {2019}
}

@article{Parrinello1984,
  author  = {Parrinello, M. and Rahman, A.},
  journal = {J. Chem. Phys.},
  volume  = {80},
  pages   = {860--867},
  year    = {1984}
}

@article{hele2015boltzmann,
  title={Boltzmann-conserving classical dynamics in quantum time-correlation functions:“Matsubara dynamics”},
  author={Hele, Timothy J H and Willatt, Michael J and Muolo, Andrea and Althorpe, Stuart C},
  journal=JCP,
  volume={142},
  number={13},
  pages={134103},
  year={2015},
  publisher={AIP Publishing}
}

@article{coalson1986connection,
  title={On the connection between Fourier coefficient and Discretized Cartesian path integration},
  author={Coalson, Rob D},
  journal=JCP,
  volume={85},
  number={2},
  pages={926--936},
  year={1986},
  publisher={American Institute of Physics}
}

@article{prada2023comparison,
  title={Comparison of Matsubara dynamics with exact quantum dynamics for an oscillator coupled to a dissipative bath},
  author={Prada, Adam and P{\'o}s, Eszter S and Althorpe, Stuart C},
  journal=JCP,
  volume={158},
  number={11},
  pages={114106},
  year={2023},
  publisher={AIP Publishing}
}

@book{risken1989fokker,
  title={The Fokker-Planck Equation: Methods of Solution and Applications},
  author={Risken, Hannes},
  year={1989},
  publisher={Springer}
}

@article{Meier1999,
  author  = {Meier, C. and Tannor, D. J.},
  journal = {J. Chem. Phys.},
  volume  = {111},
  pages   = {3365--3376},
  year    = {1999}
}

@book{zwanzig2001nonequilibrium,
  title={Nonequilibrium statistical mechanics},
  author={Zwanzig, Robert},
  year={2001},
  publisher={Oxford university press},
}

@article{kubo1966fluctuation,
  title={The fluctuation-dissipation theorem},
  author={Kubo, Rep},
  journal={Rep. Prog. Phys.},
  volume={29},
  number={1},
  pages={255},
  year={1966},
  publisher={IOP Publishing}
}

@article{baczewski2013numerical,
  title={Numerical integration of the extended variable generalized Langevin equation with a positive Prony representable memory kernel},
  author={Baczewski, Andrew D and Bond, Stephen D},
  journal=JCP,
  volume={139},
  number={4},
  pages={044107},
  year={2013},
  publisher={AIP Publishing}
}

@article{ceperley1995path,
  title={Path integrals in the theory of condensed helium},
  author={Ceperley, David M},
  journal=RMP,
  volume={67},
  number={2},
  pages={279},
  year={1995},
  publisher={APS}
}

@misc{moore2026research,
  author       = {Moore, William H D},
  title        = {Research data supporting ``{E}quilibrating continuous-variable open quantum systems using stochastic classical trajectories in path-integral space''},
  year         = {2026},
  publisher    = {University of Cambridge Repository}
}

\end{document}